\documentclass[preprintnumbers,showkeys,showpacs,byrevtex]{revtex4}
\usepackage{amsmath,amsfonts,amssymb,amscd,amsxtra,amsthm}
\usepackage{graphicx}
\usepackage{epstopdf}
\usepackage{bm}
\begin{document}
\preprint{PKNU-NuHaTh-2013-04}
\preprint{KIAS-P13029}
\title{Thermal conductivity of the quark matter for the SU(2) light-flavor sector }
\author{Seung-il Nam}
\email[E-mail: ]{sinam@pknu.ac.kr; sinam@kias.re.kr}
\affiliation{Department of Physics, Pukyong National University (PKNU), Busan 608-737, Republic of Korea\\
School of Physics, Korea Institute for Advanced Study (KIAS), Seoul 130-722, Republic of Korea\\
Asia Pacific Center for Theoretical Physics (APCTP), Pohang 790-784, Republic of Korea}
\date{\today}
\begin{abstract}
We investigate the thermal conductivity ($\kappa$) of the quark matter at finite quark chemical potential  $(\mu)$ and temperature $(T)$, employing the Green-Kubo formula, for the SU(2) light-flavor sector with the finite current-quark mass $m=5$ MeV. As a theoretical framework, we construct an effective thermodynamic potential from the $(\mu,T)$-modified liquid-instanton model (mLIM). Note that all the relevant model parameters are designated as functions of $T$, using the trivial-holonomy caloron solution. By solving the self-consistent equation of mLIM, we acquire the constituent-quark mass $M_0$ as a function of $T$ and $\mu$, satisfying the universal-class patterns of the chiral phase transition. From the numerical results for $\kappa$, we observe that there emerges a peak at $\mu\approx200$ MeV for the low-$T$ region, i.e. $T\lesssim100$ MeV. As $T$ increase over $T\approx100$ MeV, the curve for $\kappa$ is almost saturated as a function of $T$ in the order of $\sim10^{-1}\,\mathrm{GeV}^2$, and grows with respect to $\mu$ smoothly. At the normal nuclear-matter density $\rho_0=0.17\,\mathrm{fm}^{-3}$, $\kappa$ shows its maximum $6.22\,\mathrm{GeV}^2$ at $T\approx10$ MeV, then decreases exponentially down to $\kappa\approx0.2\,\mathrm{GeV}^2$.  We also compute the ratio of $\kappa$ and the entropy density, i.e. $\kappa/s$ as a function of $(\mu,T)$ which is a monotonically decreasing function for a wide range of $T$, then approaches a lower bound  at very high $T$: $\kappa/s_\mathrm{min}\gtrsim0.3\,\mathrm{GeV}^{-1}$ in the vicinity of $\mu=0$. 
\end{abstract}
\maketitle
\section{Introduction}
The transport coefficients $(\mathcal{C}_T)$, such as the shear $(\eta)$ and bulk $(\zeta)$ viscosities, and electrical $(\sigma)$ and thermal $(\kappa)$ conductivities, characterize thermodynamic properties of a matter. When a thermodynamic system starts deviating from its equilibrium by a thermodynamic force $(F)$, a flux $(\Phi)$ appears to make the system equilibrated. These two quantities, the force and flux, are related with the transport coefficients in a linear form, i.e. $\Phi=\mathcal{C}_T F$. It has been well known that the transport coefficients can be theoretically described by the Green-Kubo formula in terms of the linear response theory~\cite{Green-Kubo:1957mj}. Recent developments in the heavy-ion collision (HIC) experiments at the Relativistic Heavy-Ion Collider (RHIC) of Brookhaven National Laboratory (BNL) and Large Hadron Collider (LHC) of European Organization for Nuclear Research (CERN) have revealed various interesting features of the Quark-Gluon Plasma (QGP), which has been taken into account as an equilibrated thermodynamic system. The existence of QGP was explored intensively, and many experimental and theoretical observations have supported its existence. One of the most interesting findings for QGP must be that it behaves like a nearly perfect fluid, being indicated by the small value of the shear viscosity~\cite{Policastro:2001yc,Buchel:2003tz,Arsene:2004fa}. Note that the lower bound for $\eta/s$, which was estimated by the AdS/QCD calculations and called as the KSS bound $\approx1/4\pi$ ~\cite{Kovtun:2004de}, matches well with the experimental data eventually, in comparison with the hydrodynamic simulations~\cite{Song:2010mg}. 

Among the transport coefficients, we would like to study the thermal conductivity $(\kappa)$ of the quark matter at finite temperature $(T)$ and quark chemical potential $(\mu)$. To date, there have been many theoretical works for $\kappa$: The kinetic theory approach for the strongly interacting systems~\cite{Prakash:1993bt}, chiral perturbation theory (ChPT) for the thermodynamic system composed of the meson gas ~\cite{FernandezFraile:2009mi}, Chapman-Enskog approximation in ultra-relativistic limit~\cite{Mattiello:2012yt}, Nambu--Jona-Lasinio (NJL) model~\cite{Iwasaki:2009iw,Marty:2013ita}, Nambu-Goldstone boson model for the color-flavor locked (CFL) phase~\cite{Shovkovy:2002kv,Braby:2009dw}, and so on. We note that, interestingly, different theoretical frameworks usually provide considerably distinctive results especially for the estimated strengths of $\kappa$: The NJL model with the Breit-Wigner type quark spectral density showed an almost flat but slightly decreasing curve for $\kappa$ with respect to $T=(0.15\sim0.23)$ MeV for $\mu=0$, and it was a smoothly decreasing function of $\mu=(0\sim0.3)$ GeV for $T=200$ MeV~\cite{Iwasaki:2009iw}: A characteristic strength for $\kappa$ turned out to be about $0.01\,\mathrm{GeV}^2$. In contrast, in Ref. ~\cite{Marty:2013ita}, the NJL model for $N_f=3$ estimated $\kappa$ in the order of a few $\mathrm{GeV}^2$ for $T=(0.5\sim2.0)$ GeV at $\mu=0$, showing a structural but almost flat curve with respect to $T$. In the meson (pion) gas model in terms of the chiral perturbation theory (ChPT)~\cite{FernandezFraile:2009mi}, $\kappa$ showed a similar strength for $T=(0\sim0.2)$ GeV to that of Ref.~\cite{Iwasaki:2009iw}, whereas the curve shape was quite different. Using the first Chapman-Enskog approximation, $\kappa$ was studied in ultra-relativistic limit, resulting in a increasing function of $T$ in the order of $\sim10^{-1}\,\mathrm{GeV}^2$ for $T\gtrsim T_c$~\cite{Mattiello:2012yt}. In Ref.~\cite{Kapusta:2012zb}, taking into account the mode-coupling theory, they designated a model for $\kappa$ near the critical point. Ignoring the smooth background conductivity and focusing on the region below the critical baryon density, $\kappa$ was estimated in the similar order to that of Ref.~\cite{Mattiello:2012yt} for $T\gtrsim T_c$.   

In the present work, as for the theoretical framework for computing $\kappa$, we will make use of the liquid-instanton model (LIM) for the SU(2) light-flavor sector with the finite current-quark mass $m_{u,d}\equiv m=5$ MeV, due to the isospin symmetry, throughout the present work. Then, we modify it as a function of $T$ as well $\mu$ (mLIM). In this model, the quarks acquire the dynamically generated effective mass via the non-trivial quark-instanton interactions, demonstrating the microscopic mechanism for the spontaneous breakdown of chiral symmetry (SBCS)~\cite{Schafer:1996wv,Diakonov:2002fq}. All the relevant phenomenological model parameters, such as the average (anti)instanton size $\bar{\rho}$ and inter-(anti)instanton distance $\bar{R}$, are given as functions of $T$, using the trivial-holonomy caloron solution~\cite{Harrington:1976dj,Diakonov:1988my}. In doing that, we assume that the caloron distribution is not affected by $\mu$ for simplicity, since it is derived basically from the Yang-Mills equation in Euclidean space. Considering all the ingredients mentioned above, we are ready to construct an effective thermodynamic potential in order for exploring the chiral phase-transition diagram as a function of $(T,\mu)$. By solving the self-consistent (gap) equation with respect to the constituent-quark mass at the zero virtuality $Q^2=0$, $M_0$, we observe that a crossover chiral phase-transition line turns into the first order one at $(\mu,T)=(235,56)$, i.e. the critical end point (CEP). Moreover, the critical values for chiral transitions turn out to be $(\mu_c,T_c)\approx(342,156)$ MeV within mLIM. Being equipped with the ($\mu$-$T$) dependent quark mass $M_0$, which satisfies the proper chiral phase transition pattern for the finite SU(2) current-quark mass, the Green-Kubo formula is used for computing $\kappa$, following closely the procedure as in Ref.~\cite{Iwasaki:2009iw}. As already applied to the computation for the shear viscosity $\eta$ in our previous work~\cite{Nam:2013fpa}, we again make use of the Gaussian-type quark spectral density, motivated by the instanton physics. 

From the numerical calculations, as for the low-$T$ region, i.e. $T\lesssim100$ MeV, the $\mu$ dependence of $\kappa$ curve becomes nontrivial: At a fixed value of $T$, the curve of $\kappa$ increases until $\mu\approx200$ MeV, then decreases exponentially, due to that the derivative of the Fermi distribution and the quark spectral density give their maximum near the Fermi energy. It turns out that, in the high-$T$ region above $\sim100$ MeV, the curve of $\kappa$ is almost flat as a function of $T$, and it looks like that of the meson-gas model~\cite{FernandezFraile:2009mi} qualitatively, although the obtained order of the strength is $\sim10^{-1}\,\mathrm{GeV}^2$, being similar to those from Refs.~\cite{Mattiello:2012yt,Kapusta:2012zb}. Moreover, $\kappa$ increases smoothly and quadratically with respect to $\mu$ at a fixed $T$ value. For instance, at $T=200$ MeV, $\kappa$ becomes about $0.13\,(0.27)$ for $\mu=0\,(300)$ MeV. This observation is obviously in contrast to that of that of Ref.~\cite{Iwasaki:2009iw}.  From these observations, we can conclude that the high-$T$ quark matter, possibly QGP, is a good thermal conductor in comparison to usual matters.  At the normal nuclear matter density $\rho=0.17\,\mathrm{fm}^{-3}$, in the low-$T$ region, $\kappa$ increases up to $6.22\,\mathrm{GeV}^2$ at $T\approx10$ MeV, then decreases into the order of $10^{-1}\,\mathrm{GeV}^2$, continuing to that in the high-$T$ region. We also compute the ratio of $\kappa$ and the entropy density, i.e. $\kappa/s$ as a function of $(\mu,T)$, and observe that it is a monotonically decreasing function for a wide range of $T$. Being similar to the ratio of the shear viscosity and the entropy density $\eta/s$ for QGP, a lower-bound value for $\kappa/s$ is observed at very high $T\approx0.85$ GeV: $\kappa/s\gtrsim0.3\,\mathrm{GeV}^{-1}$ in the vicinity of $\mu=0$. 

The present work is structured as follows: In Section II, we introduce the theoretical framework for computing $\kappa$: The Green-Kubo formula for the transport coefficients, the caloron solution, and the effective thermodynamic potential of mLIM in a consistent way. The numerical results for $\kappa$ and relevant discussions are given in Section III. Section IV is devoted to the summary and future perspectives.

\section{Thermal conductivity of quark matter at finite $\mu$ and $T$}
\subsection{Thermal conductivity via the Green-Kubo formula}
In this Subsection, we briefly introduce the Green-Kubo formula for the thermal conductivity $\kappa$, and derive an analytic expression for $\kappa$ using the Gaussian-type quark spectral density, which was developed for the shear viscosity $\eta$ in the previous work~\cite{Nam:2013fpa}. According to the Green-Kubo formula, the (static) thermal conductivity $\kappa$ can be defined in Minkowski space as follows~\cite{Iwasaki:2009iw}: 
\begin{equation}
\label{eq:DEFKAPPA}
\kappa=\lim_{\omega\to0}\kappa(\omega)=-\frac{1}{3T}
\frac{d}{d\omega}\mathrm{Im}\Pi^R(\omega)\vert_{\omega=0}
\end{equation}
where $\omega$ represents the frequency of a thermodynamic system. $\Pi^R$ stands for the retarded Green's function and can be related to that with an imaginary Matsubara frequency $\omega_n=2n\pi T$:
\begin{equation}
\label{eq:RG}
\Pi^R(\omega)=\Pi(i\omega_n)|_{i\omega_n=\omega+i\epsilon}.
\end{equation}

Closely following the derivation of Ref.~\cite{Iwasaki:2009iw}, the analytic expression for $\kappa$ as a function of $(\mu,T)$ can be written in terms of the quark spectral density $\rho_k$ as follows:
\begin{eqnarray}
\label{eq:KAPPATMU}
\kappa(T,\mu)&=&-\frac{1}{12T}\int\frac{d^3\bm{k}}{(2\pi)^3}\frac{dk_0}{2\pi}
\frac{\partial n(k_0)}{\partial k_0}\mathrm{Tr}_{c,f,\gamma}
\left[H_{\bm{k},\mu}\,\rho_{k,\mu}\,\gamma_i\,H_{\bm{k},\mu}\,\rho_{k,\mu}\,\gamma_i
+H_{\bm{k},\mu}^2\,\rho_{k,\mu}\,\gamma_i\,\rho_{k,\mu}\,\gamma_i\right],
\end{eqnarray}
where we have used a simplified notation that $H_{\bm{k},\mu}\equiv\gamma_0(\bm{k}\cdot\bm{\gamma}+M_{\bm{k}})-\mu$, and $\gamma_i$ denotes the $i$-th component of the gamma matrix. The Fermi distribution is denoted by $n(k_0)=1/(1+e^{k_0/T})$. In the generic liquid-instanton model (LIM), the quarks obtain the momentum-dependent effective mass through the nontrivial interactions with the (anti)instantons at the zero mode~\cite{Diakonov:2002fq}. Being motivated by this mechanism, we introduce a parametrization of the mass as~\cite{Nam:2012sg}:
\begin{equation}
\label{eq:EQM}
M_{\bm{k}}=M_0(T,\mu)\left[\frac{2}{2+\bar{\rho}^2(T)\,\bm{k}^2} \right]^3.
\end{equation}
Note that the constituent quark mass at zero virtuality $M_0$ and the average instanton size $\bar{\rho }$ are functions of $T$ and/or $\mu$ implicitly here. We will discuss them in detail in the next Subsection. As suggested for the shear viscosity for quark matter in Ref.~\cite{Nam:2012sg}, we take into account the quark spectral density with a finite width, which relates to the model scale, i.e the average instanton size $\bar{\rho}$: $\Lambda\sim1/\bar{\rho}$. Then, the quark spectral density $\rho_{k,\mu}$ reads as a function of $(k_0,\bm{k},\mu)$ explicitly and $T$ implicitly.
\begin{equation}
\label{eq:RHOREG}
\rho_{k,\mu}\equiv\rho_{k,\mu}(k_0,\bm{k},\mu)\approx2\pi\,\mathrm{sgn}[k_0]\,[\gamma_0k_0-\bm{\gamma}\cdot\bm{k}+\bar{M}_{\bm{k}}]\mathcal{F}_{k,\mu},
\end{equation}
where we defined $\bar{M}_{\bm k}\equiv{M}_{\bm k}+m$ and the Gaussian-type distribution is defined as
\begin{equation}
\label{eq:FW}
\mathcal{F}_{k,\mu}=\frac{1}{2\sqrt{2\pi}E_{\bm{k}}\Lambda}\left[\exp\left[-\frac{(k_0+\mu-E_{\bm{k}})^2}{2\Lambda^2} \right]+\exp\left[-\frac{(k_0+\mu+E_{\bm{k}})^2}{2\Lambda^2} \right] \right]
\end{equation}
with the quark energy $E^2_{\bm{k}}=\bm{k}^2+\bar{M}^2_{\bm{k}}$. It is worth mentioning that the quark spectral density in Eq.~(\ref{eq:RHOREG}) satisfies its normalization condition, $\int\rho_{k,\mu} dk_0=2\pi\gamma_0$ for vacuum. Note that the quark spectral density has maxima at $k_0=\pm E_{\bm{k}}-\mu$ as expected. We also verified that Eq.~(\ref{eq:RHOREG}), combining with Eq.~(\ref{eq:EQM}), gives the numerical result for the quark condensate for vacuum $\langle\bar{q}q\rangle\approx-(250\,\mathrm{MeV})^3$ with $\bar{\rho}\approx1/3$ fm and $M_0(0,0)\approx350$ MeV which is well compatible with its empirical one~\cite{Nam:2012sg}. The trace in Eq.~(\ref{eq:KAPPATMU}) can be performed straightforwardly, resulting in
\begin{eqnarray}
\label{eq:KAPPATMU1}
\kappa(T,\mu)&=&\frac{N_cN_f}{6\pi^2T^2}
\int\,d^3\bm{k}\,dk_0\,
n(k_0)[1-n(k_0)]\mathcal{F}^2_{k,\mu}
\left[(k^2_0+E^2_{\bm{k}}-4k_0\mu+2\mu^2)k^2_i+\mu^2(k^2_0-E^2_{\bm{k}})\right]
\cr
&=&
\frac{2N_cN_f}{3\pi T^2}
\int\,\mathrm{k}^2d\mathrm{k}\,dk_0\,
n(k_0)[1-n(k_0)]\mathcal{F}^2_{k,\mu}\left[(k^2_0+E^2_{\bm{k}}-4\mu k_0+2\mu^2)\frac{\mathrm{k}^2}{3}+\mu^2(k^2_0-E^2_{\bm{k}})\right],
\end{eqnarray}
where we have used a notation $\mathrm{k}\equiv|\bm{k}|$ for convenience, and considered in the second line of Eq.~(\ref{eq:KAPPATMU1}) that $\int d^3\bm{k}f(\bm{k}^2)k^2_i=\frac{4\pi}{3}\int\,\mathrm{k}^4f(\mathrm{k}^2)d\mathrm{k}$, due to the rotational symmetry of the system.
\subsection{Instantons at finite temperature: Caloron solution}
Now, we are in a position to explain how to modify the instanton parameters as functions of $T$. As in Ref.~\cite{Nam:2009nn}, we make use of the caloron distribution with the trivial holonomy, i.e. Harrington-Shepard caloron for this purpose~\cite{Harrington:1976dj,Diakonov:1988my}. An instanton distribution function for arbitrary $N_c$ and $N_f$ can be written with a Gaussian suppression factor as a function of $T$ and an arbitrary instanton size $\rho$ for the pure-glue QCD, i.e. Yang-Mills equation in Euclidean space~\cite{Diakonov:1988my}:
\begin{equation}
\label{eq:IND}
d(\rho,T)=\underbrace{C_{N_c}\,\Lambda^b_{\mathrm{RS}}\,
\hat{\beta}^{N_c}}_\mathcal{C}\,\rho^{b-5}
\exp\left[-(A_{N_c}T^2
+\bar{\beta}\gamma n\bar{\rho}^2)\rho^2 \right].
\end{equation}
Note that the $CP$-invariant vacuum was taken into account in Eq.~(\ref{eq:IND}), and we assumed the same analytical form of the distribution function for the (anti)instanton. The instanton number density (packing fraction) $N/V\equiv n\equiv1/\bar{R}^4$ and $\bar{\rho}$ have been taken into account as functions of $T$ implicitly here. To make the problems easy, we choose that the numbers of the anti-instanton and instanton are the same, i.e. $N_I=N_{\bar{I}}=N$. We also assigned the constant factor in the right-hand-side of the above equation as $\mathcal{C}$ for simplicity. The abbreviated notations are given as follows:
\begin{eqnarray}
\label{eq:PARA}
\hat{\beta}&=&-b\ln[\Lambda_\mathrm{RS}\rho_\mathrm{cut}],\,\,\,\,
\bar{\beta}=-b\ln[\Lambda_\mathrm{RS}\langle R\rangle],\,\,\,
C_{N_c}=\frac{4.60\,e^{-1.68\alpha_{\mathrm{RS}} Nc}}{\pi^2(N_c-2)!(N_c-1)!},
\cr
A_{N_c}&=&\frac{1}{3}\left[\frac{11}{6}N_c-1\right]\pi^2,\,\,\,\,
\gamma=\frac{27}{4}\left[\frac{N_c}{N^2_c-1}\right]\pi^2,\,\,\,\,
b=\frac{11N_c-2N_f}{3}.
\end{eqnarray}
Note that we defined the one-loop inverse charges $\hat{\beta}$ and $\bar{\beta}$ at certain phenomenological cutoff value $\rho_\mathrm{cut}$ and $\langle R\rangle\approx\bar{R}$. $\Lambda_{\mathrm{RS}}$ stands for a scale, depending on a renormalization scheme, while $V_3$ for the three-dimensional volume. Using the instanton distribution function in Eq.~(\ref{eq:IND}), we can compute the average value of the instanton size $\bar{\rho}^2$ straightforwardly as follows~\cite{Schafer:1996wv}:
\begin{equation}
\label{eq:rho}
\bar{\rho}^2(T)
=\frac{\int d\rho\,\rho^2 d(\rho,T)}{\int d\rho\,d(\rho,T)}
=\frac{\left[A^2_{N_c}T^4
+4\nu\bar{\beta}\gamma n \right]^{\frac{1}{2}}
-A_{N_c}T^2}{2\bar{\beta}\gamma n},
\end{equation}
where $\nu=(b-4)/2$. It is clear that Eq.~(\ref{eq:rho}) satisfies the  following asymptotic behaviors~\cite{Schafer:1996wv}:
\begin{equation}
\label{eq:asym}
\lim_{T\to0}\bar{\rho}^2(T)=\sqrt{\frac{\nu}{\bar{\beta}\gamma n}},
\,\,\,\,
\lim_{T\to\infty}\bar{\rho}^2(T)=\frac{\nu}{A_{N_c}T^2}.
\end{equation}
Here, the second relation of Eq.~(\ref{eq:asym}) shows a correct scale-temperature behavior at high $T$, i.e., $1/\bar{\rho}\approx\Lambda\propto T$. Substituting Eq.~(\ref{eq:rho}) into Eq.~(\ref{eq:IND}), the caloron distribution function can be evaluated further:
\begin{equation}
\label{eq:dT}
d(\rho,T)=\mathcal{C}\,\rho^{b-5}
\exp\left[-\mathcal{F}(T)\rho^2 \right],\,\,\,\,
\mathcal{F}(T)=\frac{1}{2}A_{N_c}T^2+\left[\frac{1}{4}A^2_{N_c}T^4
+\nu\bar{\beta}\gamma n \right]^{\frac{1}{2}}.
\end{equation}
The instanton packing fraction $n$ can be computed self-consistently, using the following equation:
\begin{equation}
\label{eq:NOVV}
n^\frac{1}{\nu}\mathcal{F}(T)=\left[\mathcal{C}\,\Gamma(\nu) \right]^\frac{1}{\nu},
\end{equation}
where we replaced $NT/V_3\to n$, and $\Gamma(\nu)$ stands for the $\Gamma$-function with an argument $\nu$. Note that $\mathcal{C}$ and $\bar{\beta}$ can be determined easily using Eqs.~(\ref{eq:rho}) and (\ref{eq:NOVV}), incorporating the vacuum values for $n\approx(200\,\mathrm{MeV})^4$ and $\bar{\rho}\approx(600\,\mathrm{MeV})^{-1}$: $\mathcal{C}\approx9.81\times10^{-4}$ and $\bar{\beta}\approx9.19$. Finally, in order for estimating the $T$-dependence of $M_0$, it is necessary to consider the normalized distribution function, defined as follows,
\begin{equation}
\label{eq:NID}
d_N(\rho,T)=\frac{d(\rho,T)}{\int d\rho\,d(\rho,T)}
=\frac{\rho^{b-5}\mathcal{F}^\nu(T)
\exp\left[-\mathcal{F}(T)\rho^2 \right]}{\Gamma(\nu)}.
\end{equation}
Here, the subscript $N$ denotes the normalized distribution. For brevity, we want to employ the large-$N_c$ limit to simplify the expression for $d_N(\rho,T)$. In this limit, as understood from Eq.~(\ref{eq:NID}), $d_N(\rho,T)$ can be approximated as a $\delta$-function:
\begin{equation}
\label{eq:NID2}
\lim_{N_c\to\infty}d_N(\rho,T)=\delta[{\rho-\bar{\rho}(T)}].
\end{equation}

\begin{figure}[t]
\begin{tabular}{cc}
\includegraphics[width=8.5cm]{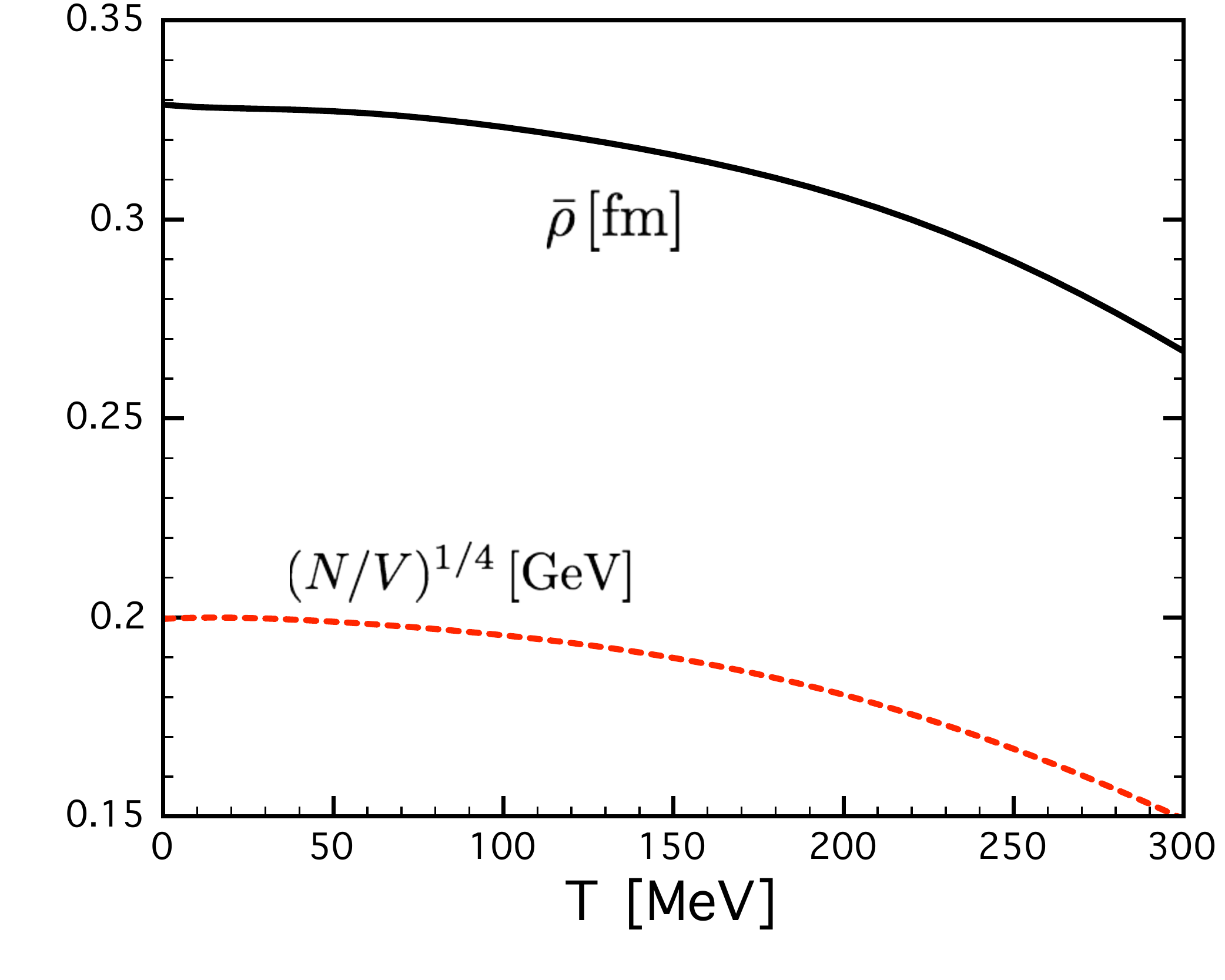}
\includegraphics[width=10cm]{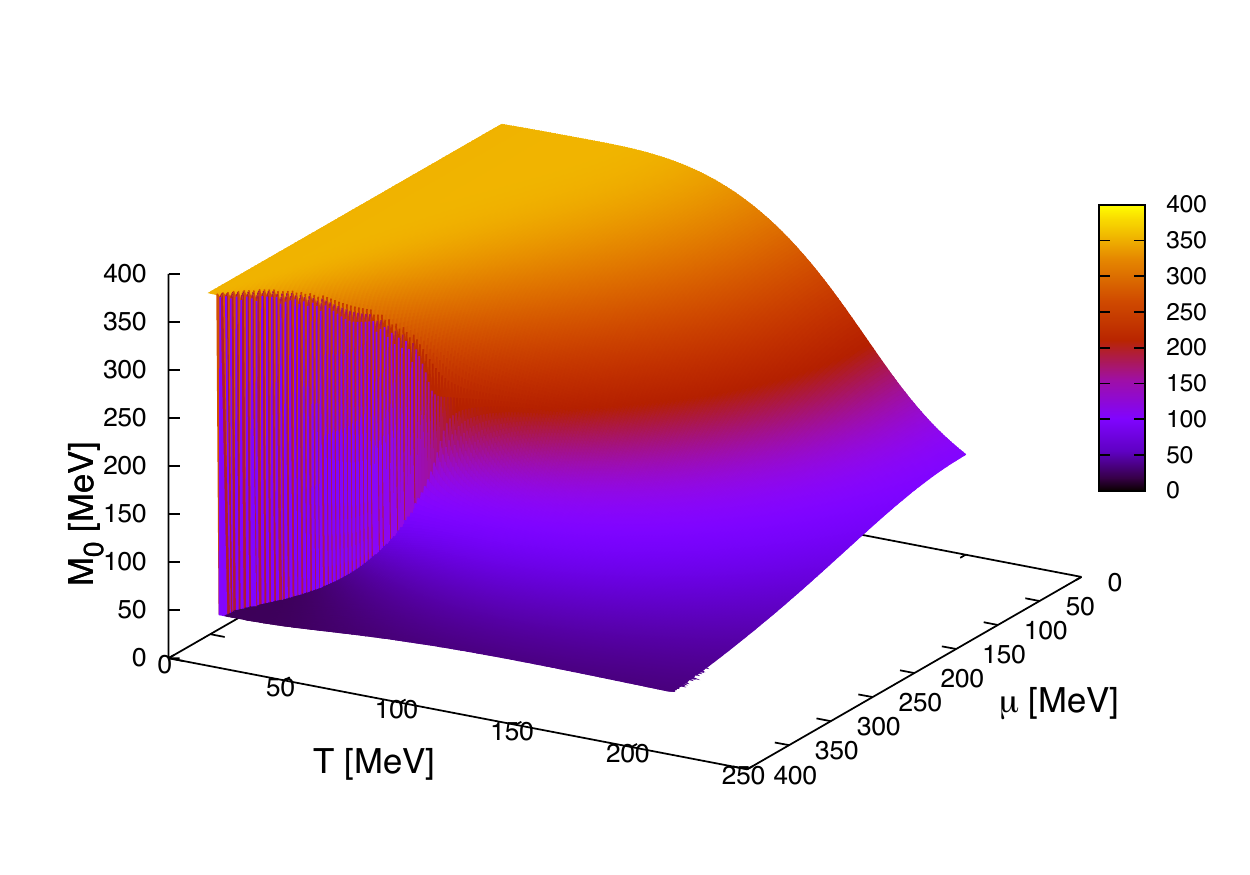}
\end{tabular}
\caption{(Color online) Left: Temperature dependence of the instanton parameters: average (anti)instanton size $\bar{\rho}(T)$ (solid) and (anti)instanton number density (packing fraction) $N/V$ (dash) as functions of $T$. The decreasing behaviors of the curves indicate the partial restoration of SB$\chi$S. Right: QCD phase diagram via the constituent-quark mass $M_0$ as a function of $T$ and $\mu$ for $m=5$ MeV by solving Eq.~(\ref{eq:LIMGAP}). The color scale indicates the strength of $M_0$. We find that the critical end point (CEP) locates at $(T,\mu)\approx(56,235)$ MeV. }       
\label{FIG12}
\end{figure}

The numerical results for $\bar{\rho}(T)$  and $N/V(T)$ are given in the left panel of Figure~\ref{FIG12}. The curve for $\bar{\rho}(T)$ indicates that the average (anti)instanton size smoothly decreases as a function of $T$. This tendency shows that the instanton ensemble becomes diluted and the nonperturbative effects from the quark-instanton interactions are reduced. At $T=(150\sim200)$ MeV, which is close to the chiral phase transition $T$, the instanton size gets reduced by about $(10\sim20)\%$ in comparison to its value at $T$=0. Taking into account that the instanton size corresponds to the scale parameter of the present model, i.e. UV cutoff mass, $\bar{\rho}\approx1/\Lambda$, the $T$-dependent cutoff mass is a clearly distinctive feature in comparison to other low-energy effective models, such as the NJL model. Similarly, the instanton number density $N/V$ decreases as $T$ increases: The instanton ensemble becomes diluted with respect to $T$. 

\subsection{Effective thermal dynamic potential at finite $T$ and $\mu$}
Since all the thermodynamic properties can be studied via a thermodynamic potential, we first construct an effective one using all the ingredients discussed in the previous Subsections. As in Ref.~\cite{Nam:2009nn}, the mLIM thermodynamic potential per volume in the leading large-$N_c$ contributions for finite $(\mu,T)$ can be written as follows:
\begin{eqnarray}
\label{eq:ETP}
\Omega_\mathrm{mLIM}
&=&\mathcal{C}+\frac{N}{V}\ln\lambda+2\sigma^2-2N_cN_f\int_{\bm{k}}
\left\{E_{\bm{k}}+T\ln\left[F_+F_- \right]\right\},\,\,\,\,
F_{\pm}\equiv1+e^{-\frac{E_{\bm{k}}\pm\mu}{T}}  
\end{eqnarray}
where $\int_{\bm{k}}\equiv\int\frac{d^3\bm{k}}{(2\pi)^3}$ for brevity. $\lambda$ and $\mathrm{M}$ represent a Lagrange multiplier to exponentiate the effective quark-instanton action and an arbitrary mass parameter to make the argument for the logarithm dimensionless. $\sigma$ stands for the isosinglet scalar meson field corresponding to the effective quark mass. In the leading large-$N_c$ contributions, we have the relation $2\sigma^2=N/V$~\cite{Nam:2009nn}. Then, the saddle-point equation can be derived from Eq.~(\ref{eq:ETP}) by differentiating $\Omega_\mathrm{mLIM}$ by the Lagrange multiplier $\lambda$:
\begin{equation}
\label{eq:LIMGAP}
\frac{\partial\Omega_\mathrm{mLIM}}{\partial \lambda}=0\to
\frac{N}{V}-2N_c\int_{\bm{k}}
\frac{M_{\bm{k}}\bar{M}_{\bm{k}}}{E_{\bm{k}}}\left[1+\frac{1-F_-}{F_-}+\frac{1-F_+}{F_+} \right]=0
\end{equation}
Here, $\bar{\rho}=\bar{\rho}(T)$ and $M_0=M_0(T)$ implicitly. In deriving Eq.~(\ref{eq:LIMGAP}), we have used that $\partial M_0/\partial\lambda=M_0/(2\sqrt{\lambda})$, where $M_0$ appears in $E_{\bm k}$~\cite{Nam:2009nn}. Note that one can write the instanton number density in terms of the effective quark mass $M_0$ and $\bar{\rho}$~\cite{Diakonov:2002fq}:
\begin{equation}
\label{eq:NOV}
\frac{N}{V}=\frac{\mathcal{C}_0N_cM^2_0}{\pi^2\bar{\rho}^2}.
\end{equation}
The value of the real-positive parameter $\mathcal{C}_0$ is determined to reproduce $M_0=350$ MeV at $(T,\mu)=0$, resulting in $\mathcal{C}_0=0.151$ by solving the saddle-point equation in Eq.~(\ref{eq:LIMGAP}). Once $\mathcal{C}_0$ fixed, by solving Eq.~(\ref{eq:LIMGAP}) with respect to $M_0$ on the $(\mu,T)$ plane numerically, we can obtain $M_0$  as a function of $(\mu,T)$, and it is given in the right panel of Figure~\ref{FIG12} for $m=5$ MeV. Here, we have assumed that the model parameters, $\bar{\rho}$ and $N/V=1/\bar{R}^4$, are only the functions of $T$ and independent on $\mu$ for simplicity, since they were derived from the pure-glue theory. This result shows the universal patterns for the chiral phase transition appropriately: There appears the crossover chiral phase transition until the phase-transition line reaches the critical end point (CEP), then it changes into the first-order one.  From those numerical results, CEP locates at $(\mu,T)\approx(235,56)$ MeV, which is compatible to various theoretical estimations from the effective models~\cite{Stephanov:2007fk}. We will make a detailed look on the QCD phase diagram in a separated work~\cite{NAMKAO}.

Since we are also interested in the ratio of $\kappa$ and the entropy density, i.e. $\kappa/s$, we derive the entropy density from the effective potential in Eq.~(\ref{eq:ETP}) using the following:
\begin{equation}
\label{eq:ENT}
s=-\frac{\partial\Omega_\mathrm{mLIM}}{\partial T},
\end{equation}
which results in from Eqs.~(\ref{eq:ETP}) and (\ref{eq:ENT}):
\begin{equation}
\label{eq:ENT1}
s\approx-\left(\frac{\partial}{\partial T}\frac{N}{V} \right)
\left[1-\ln\left(\frac{N}{V\Lambda^4} \right) \right]+2N_cN_f\int_{\bm{k}}
\left\{\ln[F_+F_-]+\frac{F_+(F_--1)(E_{\bm k}-\mu)
+F_-(F_+-1)(E_{\bm k}+\mu)}{TF_+F_-}\right\}.
\end{equation}
In Eq.~(\ref{eq:ENT1}), we assume that $2\sigma^2\approx N/V$ and $\lambda\mathrm{M}\approx\Lambda^4$ as in the leading large-$N_c$ contributions, since $\Lambda$ is the only scale parameter of the present model. As shown in the left panel of Figure~\ref{FIG12}, the (anti)instanton number density $N/V$ is a function of $T$, so that the term $\partial (N/V)/\partial T$ in the first parenthesis in right-hand-side of Eq.~(\ref{eq:ENT1}) is finite.

\section{Numerical results and Discussions}
\begin{figure}[t]
\begin{tabular}{cc}
\includegraphics[width=8.5cm]{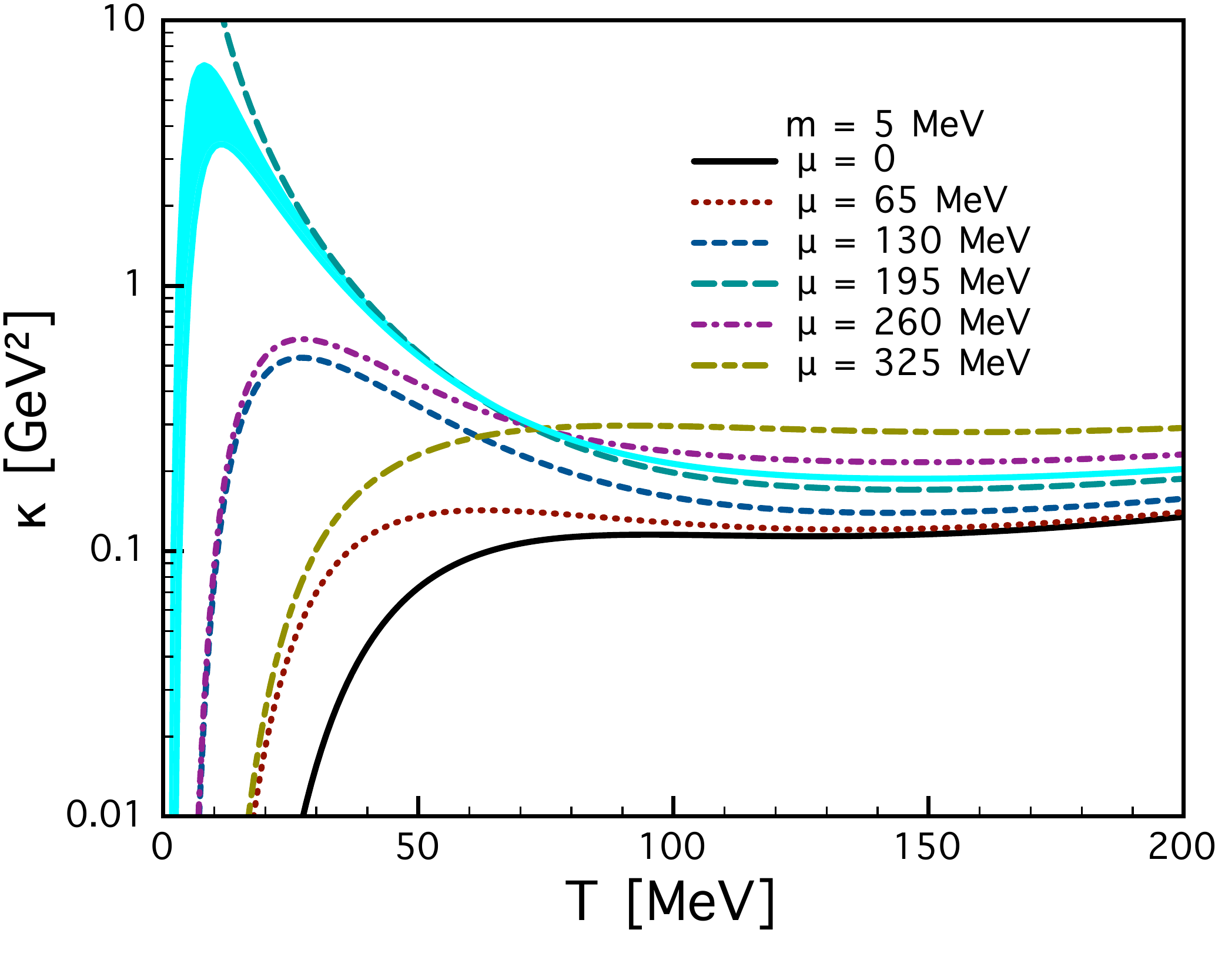}
\includegraphics[width=10cm]{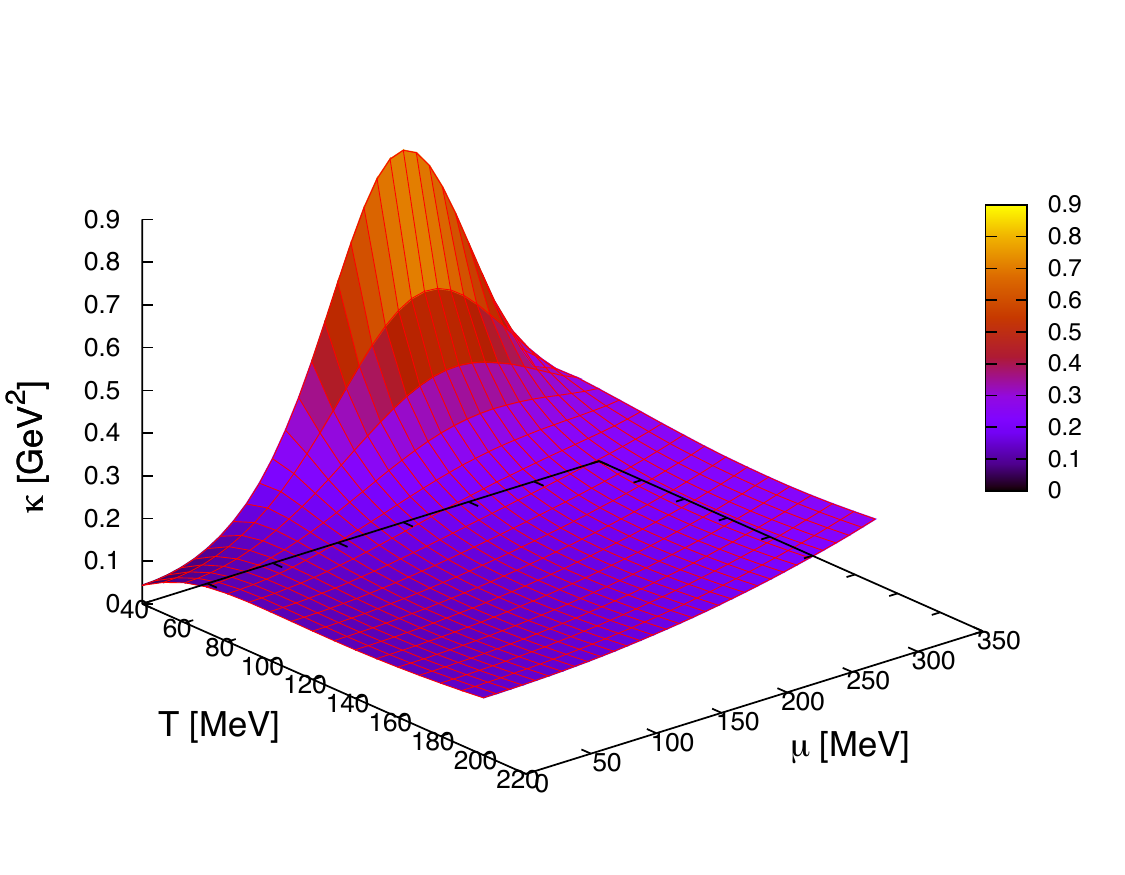}
\end{tabular}
\caption{(Color online) Left: Thermal conductivity $\kappa$ for $\mu=(0\sim325)$ MeV as a function of $T$ for the current-quark mass $m=5$ MeV. The shaded area represents $\kappa$ for the normal nuclear density $\rho_0$, corresponding to $\mu\approx(214\sim222)$ MeV as in Eq.~(\ref{eq:RHOMU1}). Right: A three-dimensional representation of $\kappa$ as a function of $(\mu,T)$ for $m=5$ MeV. The color scale bar indicates the strength of $\kappa$}       
\label{FIG34}
\end{figure}

In this Section, we present the numerical results for $\kappa$ and make relevant discussions. First, we plot the curves for $\kappa$ as functions of $T$ for the different $\mu$ values, i.e. $\mu=(0\sim325)$ MeV in the left panel of Figure~\ref{FIG34}. We observe distinctive behaviors of the curves below and above $T\approx100$ MeV: In the low-$T$ region, i.e. $T\lesssim100$ MeV, a peak appears at $\mu\approx200$ MeV for a fixed $T$ value. This situation can be clearly seen in the right panel of Figure~\ref{FIG34}, in which $\kappa$ is plotted three-dimensionally as a function of $(\mu,T)$, i.e. $\kappa$ increases up to $\mu\approx200$ MeV, then decreases smoothly. The reason for the peak is caused mainly by the derivative of the Fermi distribution as well as the quark spectral density in Eq.~(\ref{eq:KAPPATMU1}). The order of the strength of $\kappa$ varies from zero to several tens of $\mathrm{GeV}^2$, depending on $\mu$. This complicated behavior of $\kappa$ as a function of $(\mu,T)$ gets simplified for the high-$T$ region, $T\gtrsim100$ MeV. Although $\kappa$ shows a smoothly and quadratically increasing behavior with respect to $\mu$, it is almost saturated for $T$, in the order of about $10^{-1}\,\mathrm{GeV}^2$. For instance, we have $\kappa=0.12\,\mathrm{GeV}^2$ at $(\mu,T)=(0,150)$ MeV. One can understand the situation obviously by seeing the right panel of Figure~\ref{FIG34}.

Taking into account that many physical topics for the finite density are carried out at the normal nuclear-matter density $\rho_0=0.17\,\mathrm{fm}^{-3}$, it is helpful to provide theoretical information for $\kappa$ at $\rho_0$. The nuclear matter density $\rho_N$ can be related to the nucleon chemical potential $\mu^{(\rho_N)}_N$ as follows~\cite{Lee:2012fj}:
\begin{equation}
\label{eq:RHOMU1}
\rho_N=\frac{2\pi^2}{3}k^3_F,\,\,\,\,
k_F=\sqrt{[\mu^{(\rho_N)}_N]^2-[M^{(\rho_N)}_N]^2},
\end{equation}
where $\rho_N$, $k_F$, $\mu^{(\rho_N)}_N$, and $M^{(\rho_N)}_N$ indicate the nuclear-matter density, fermi momentum, nucleon chemical potential, and nucleon mass at the certain density, respectively. Considering that the nucleon mass at the normal nuclear density $\rho_N=\rho_0$ can be scaled as $M^{(\rho_0)}_N\approx(0.65\pm0.03)\,M^{(0)}_N$ at $T=0$~\cite{Larionov:2000cu}, we have $\mu^{(\rho_0)}_N\approx(641\sim667)$ MeV via Eq.~(\ref{eq:RHOMU1}). Employing a simple relation $3\mu=\mu_N$, we obtain the corresponding quark chemical potential $\mu^{(\rho_0)}\approx(214\sim222)$ MeV for the normal nuclear-matter density. In the left panel of Figure~\ref{FIG34}, the shaded area represents the possible region for $\kappa$ as a function of $T$ at $\rho_0$. We note that $\kappa$ has its maximum $6.22\,\mathrm{GeV}^2$ at $T\approx10$ MeV, then decreases exponentially down to $\sim0.2\,\mathrm{GeV}^2$ for $T=(100\sim200)$ MeV. 

Now, we want to compare out results with other theoretical estimations. In Ref.~\cite{Iwasaki:2009iw}, the authors computed $\kappa$ as a function of $(\mu,T)$. As for the quark spectral density, they employed the Breit-Wigner type one. For the rage of $T=(150\sim230)$ MeV at $\mu=0$, the curve for $\kappa$ is an almost flat but smoothly decreasing function. The strength of $\kappa$ for this $T$ region turns out to be about $10^{-2}\,\mathrm{GeV}^2$, which is about a factor-one smaller than ours. Moreover, the computed $\kappa$ decreases with respect to $\mu$ at $T=200$ MeV, whereas ours behaves contrarily. In Ref. ~\cite{Marty:2013ita}, using the NJL model for $N_f=3$,  the numerical result for $\kappa$ is in the order of a few $\mathrm{GeV}^2$ for $T=(50\sim200)$ MeV at $\mu=0$, showing a structural but almost flat curve with respect to $T$: $\kappa\approx4.4\,\mathrm{GeV}^2$ for $(\mu,T)=(0,200)$ MeV. Note that this value of a few $\mathrm{GeV}^2$ for $\kappa$ is about ten-times larger than ours. In the meson (pion) gas model in terms of the chiral perturbation theory (ChPT)~\cite{FernandezFraile:2009mi}, $\kappa$ presents a similar strength for $T=(0\sim0.2)$ GeV to that of Ref.~\cite{Iwasaki:2009iw}, whereas the curve shape was quite different. 

\begin{figure}[t]
\begin{tabular}{cc}
\includegraphics[width=8.5cm]{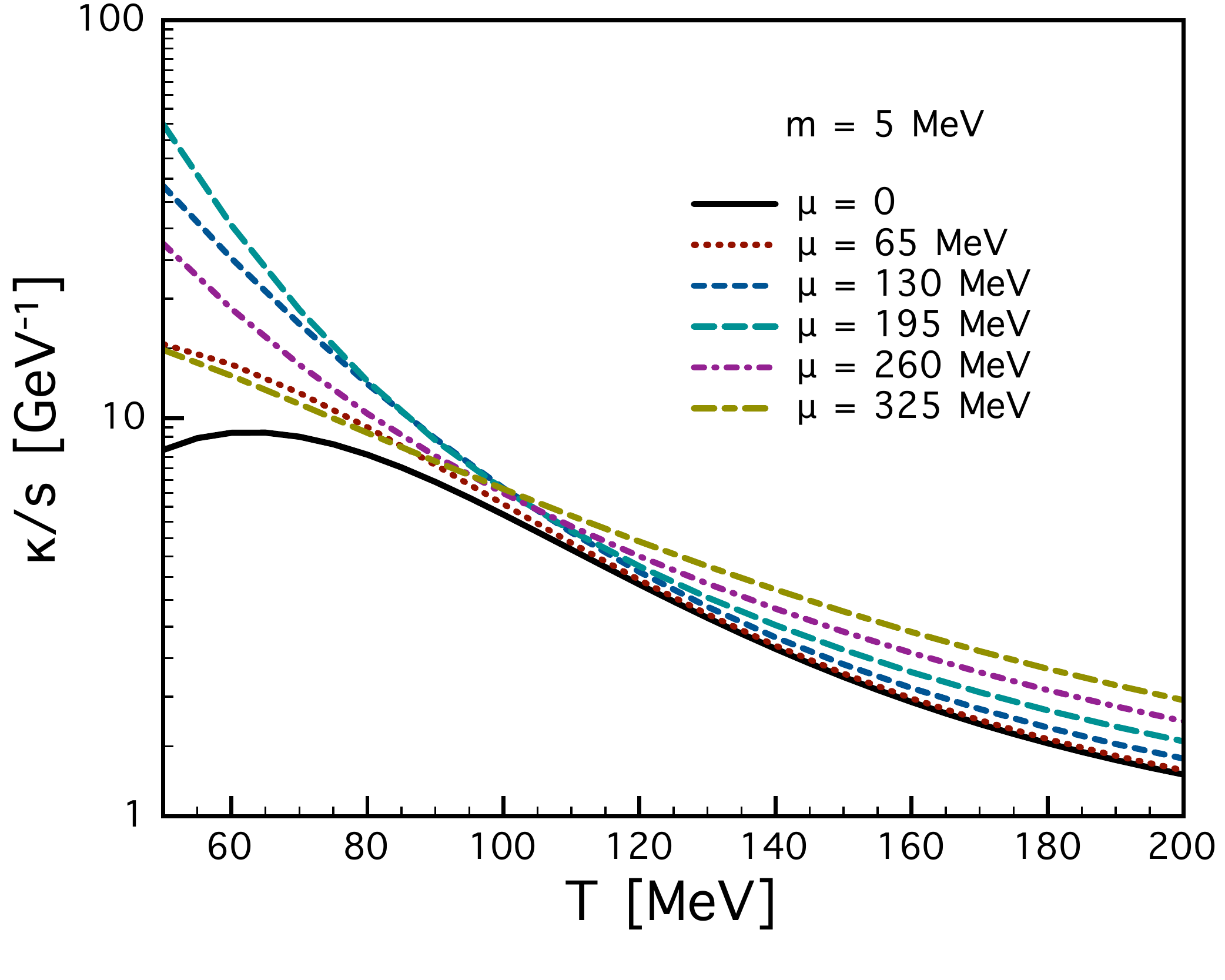}
\includegraphics[width=8.5cm]{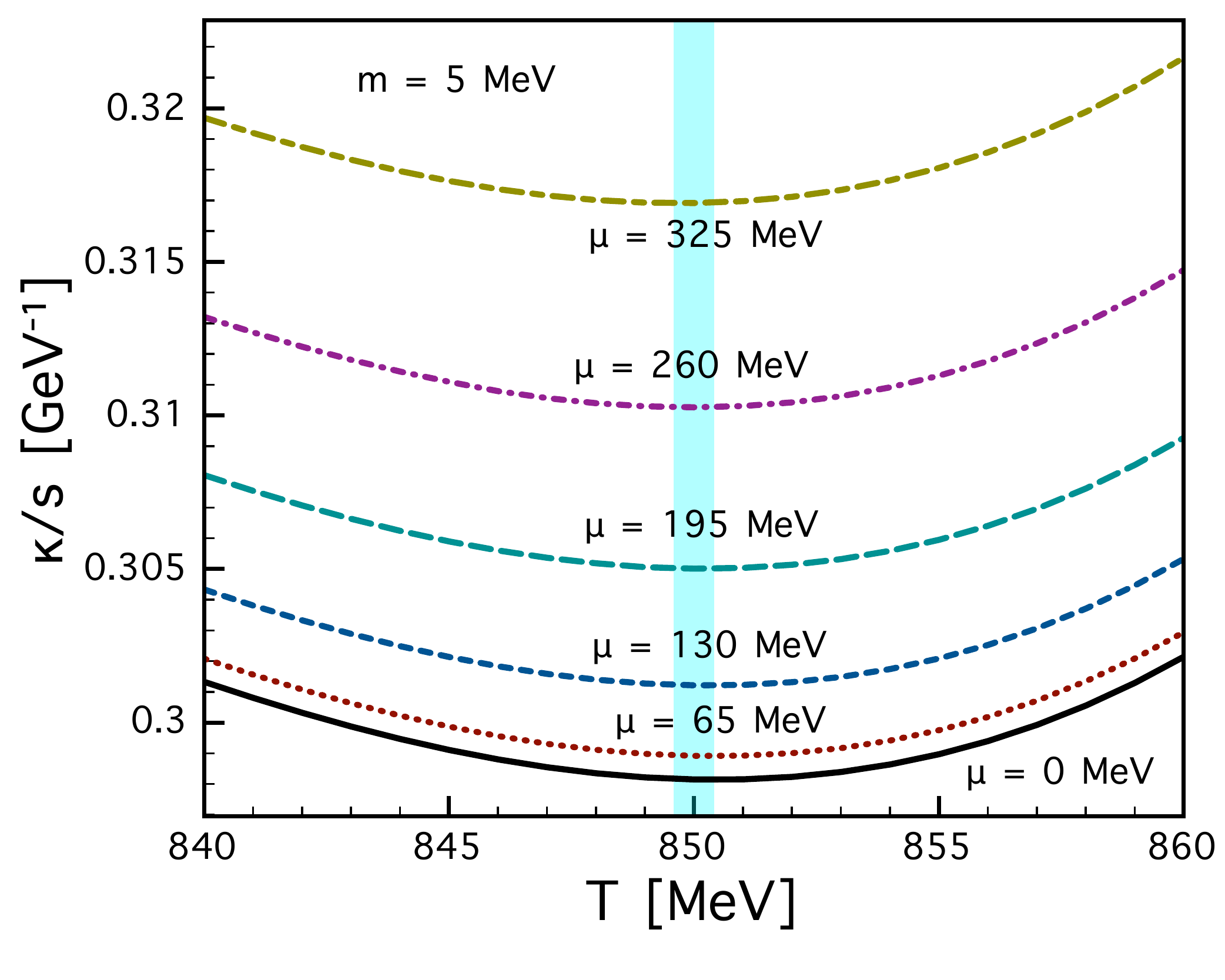}
\end{tabular}
\caption{(Color online) Left: Ratio of the thermal conductivity and entropy density, $\kappa/s$ as a function of $T=(0\sim200)$ MeV for different $\mu$ values at $m=5$ MeV. Right: $\kappa/s$ for $T=(0.84\sim0.86)$ GeV as in the left panel.}       
\label{FIG56}
\end{figure}

Finally, we present the numerical results for the ratio of $\kappa$ and the entropy density, $\kappa/s$ as a function of $T$ for the different $\mu$ values  at $m=5$ MeV in the left panel of Figure~\ref{FIG56}. Note that $\kappa/s$ decreases in general for $T\gtrsim60$ MeV for all the $\mu$ values. We also observe numerically that, if $T$ goes beyond $T\approx0.85$ GeV, the curves start to increase: Some curves for $\kappa/s$ for different $\mu$ values are shown for the vicinity of $T=0.85$ GeV in the right panel of Figure~\ref{FIG56}. Thus, $\kappa/s$ has its minimum value as a lower bound: $\kappa/s_\mathrm{min}=(0.298\sim0.317)\,\mathrm{GeV}^{-1}$ for $\mu=(0\sim325)$ MeV. Note that this observation for the lower bound is similar to the KSS bound for the ratio of the shear viscosity and the entropy density $\eta/s\approx1/(4\pi)$~\cite{Kovtun:2004de,Song:2010mg}. The value for the lower bound $\kappa/s\gtrsim0.3\,\mathrm{GeV}^{-1}$ in the vicinity of $\mu=0$ is worth being tested further by other theoretical models, such as the AdS/QCD, lattice QCD, and effective approaches.

\section{Summary}
We have studied the thermal conductivity $\kappa$ for the quark matter as a function of $T$ and $\mu$ for the SU(2) light-flavor sector: $m_{u,d}\equiv m=5$ MeV considering the isospin symmetry. The $(\mu,T)$-modified liquid-instanton model, mLIM was employed for this purpose with the $T$-dependent model parameters, which were derived from the trivial-holonomy caloron solution. We first constructed the effective thermodynamic potential and explored the QCD phase structure, by solving the self-consistent (gap) equation of the model. As a result, we obtained the constituent-quark mass as a function of $T$ and $\mu$ at zero virtuality, i.e. $M_0(T,\mu)$, and it satisfied the generic chiral restoration pattern for the finite current-quark mass for the flavor SU(2). Taking into account all the ingredients mentioned above, the thermal conductivity $\kappa$ was computed as a function of $T$ and $\mu$. In what follows, we list important observations of the present work:
\begin{itemize}
\item In the low-$T$ region, i.e. $T\lesssim100$ MeV, $\kappa$ exhibits complicated structures as a function of $(\mu,T)$. At a fixed $T$ value, there appears a peak at $\mu\approx200$ MeV. This peak is caused by the spectral quark density as well as the Fermi distribution, appearing in the analytic expression for $\kappa$.
\item As for the high-$T$ region, $T\gtrsim100$ MeV, $\kappa$ behaves very differently in comparison to that in the lower $T$ region: $\kappa$ is almost saturated in the order of $10^{-1}\,\mathrm{GeV}^2$ with respect to $T$. For instance, we have $\kappa=0.12\,\mathrm{GeV}^2$ at $(\mu,T)=(0,150)$ MeV. We also observe that the strength of $\kappa$ increases smoothly and quadratically with respect to $\mu$.
\item At the normal nuclear matter density $\rho_0=0.17\,\mathrm{fm}^{-3}$, which corresponds to $\mu=(214\sim222)$ MeV, with help of the theoretical and experimental information, $\kappa$ increases rapidly near $T=0$, then decreases exponentially. The strength of $\kappa$ varies from zero to $6.22\,\mathrm{GeV}^2$ at $T\approx10$ MeV for $T\lesssim100$ MeV, whereas it is saturated to $\sim0.2\,\mathrm{GeV}^2$ for the region beyond $T\approx100$ MeV. 
\item We also compute the ratio of $\kappa$ and the entropy density, i.e. $\kappa/s$ as a function of $(\mu,T)$ which is a monotonically decreasing and increasing function with respect to $T\gtrsim100$ MeV and $\mu$, respectively, then moves toward a lower bound at very high $T\approx0.85$ GeV: $\kappa/s_\mathrm{min}=(0.298\sim0.317)\,\mathrm{GeV}^{-1}$ for $\mu=(0\sim325)$ MeV. 
\end{itemize}

The present theoretical results can be applied to various theoretical studies on the quark-matter systems at finite $T$ and/or $\mu$, such as the core of the neutron star~\cite{Glendenning:2000zz}. Moreover, the lower-bound value for $\kappa/s$ observed theoretically in the present work is worth being studied further and tested in other theoretical approaches: $\kappa/s\gtrsim0.3\,\mathrm{GeV}^{-1}$ for $\mu\approx0$. The applications of the present theoretical results for other thermodynamic systems are under investigation and will appear elsewhere. 
\section*{Acknowledgment}
The author is grateful to C. W. Kao, Y. Lim, and K. H. Woo for fruitful discussions. We acknowledge the hospitality at APCTP where part of this work was done. This work was supported by a Research Grant of Pukyong National University (2013).

\end{document}